%% file: spec_lat04.tex
\documentclass[fleqn,twoside]{article}
\usepackage{espcrc2,isolatin1}

\usepackage{graphicx}
\usepackage[figuresright]{rotating}
\usepackage{epsfig}
\usepackage{color}

\input{macros.tex}

\newcommand{\AmS}{{\protect\the\textfont2
  A\kern-.1667em\lower.5ex\hbox{M}\kern-.125emS}}

\title{
{\vspace{-3.45cm} \normalsize \hfill
\parbox{35mm}{\hfill CPT-2004/P.067\\ \phantom{x}\hfill BU-HEP 04-15 }
}\\[22mm]
Light hadron spectra and wave functions in quenched QCD with overlap quarks 
on a large lattice\thanks{Work supported in part by US DOE
grants DE-FG02-91ER40676 and DE-AC02-98CH10866, EU contracts
HPRN-CT-2000-00145 and HPRN-CT-2002-00311, and grant
HPMF-CT-2001-01468.  We thank Boston University and NCSA for use of
their supercomputer facilities.}\thanks{Combined presentations by 
J.~Howard, L.~Lellouch and C.~Rebbi at {\it Lattice 2004}, FNAL, USA.}
}

\author{F.~Berruto\address[BNL]{Department of Physics, Brookhaven
        National Laboratory, Upton NY 11973, USA},
        N.~Garron\address[CPT]{Centre de Physique Th\'eorique$^1$, Case
        907, CNRS Luminy, F-13288 Marseille Cedex 9, France},
        C.~Hoelbling\addressmark[CPT]\thanks{Present address: 
        Department of Physics, Bergische Universit\"at Wuppertal, Gausstr. 20, D-42119, Germany},
        J.~Howard\address[BU]{Department of Physics, Boston
        University, 590 Commonwealth Avenue, Boston MA 02215, USA},
        L.~Lellouch\addressmark[CPT],
        S.~Necco\addressmark[CPT],
        C.~Rebbi\addressmark[BU],
        and
        N.~Shoresh\addressmark[BU]\thanks{Now at Harvard University.}}
       
\begin{document}

\begin{abstract}
A simulation of quenched QCD with the overlap Dirac operator has been
completed using 100 Wilson gauge configurations at $\b=6$ on an $18^3
\x 64$ lattice. We present results for meson and baryon masses, meson final
state ``wave functions'' and other observables.
\vspace{1pc}
\end{abstract}

\maketitle

\footnotetext[1]{UMR 6207 du CNRS et des universit\'es 
d'Aix-Marseille I, II et du Sud Toulon-Var, affili\'ee \`a la FRUMAM.} 

\setcounter{footnote}{0}

\section{Simulation Details}
We have studied quenched QCD on an $18^3 \times 64$ lattice, with the Wilson
gauge action at $\beta=6$ and with Neuberger's overlap Dirac operator for 
lattice fermions. We used a sample of 100 gauge configurations 
separated by 10,000 upgrades of a 6-hit Metropolis 
algorithm~(acceptance $\approx 0.5$).
A previous preliminary analysis \cite{Berruto:2003rt} used the first 
60 quark propagators that were produced; results presented here include 
all 100 quark propagators.  We refer to \cite{Berruto:2003rt} and the 
references therein for more background information and details of the
overlap operator used. New results include 
the vector meson spectrum, the use of extended sink operators in 
constructing meson correlation functions, and observation of quenched 
chiral logs. Results for kaon weak matrix elements are presented in
a separate contribution \cite{lat04_wme}.

From Wilson loops, we obtained $r_0/a=5.36\pm0.11$ $(a^{-1}=2.11\pm 0.04~{\rm
GeV})$ for the Sommer scale defined by $r_0^2F(r_0)=1.65$, $r_0=0.5~{\rm fm}$.
    
We calculated overlap quark propagators for a single point
source, for all 12 color-spin combinations and quark masses $am_q=0.03, 0.04, 
0.06, 0.08, 0.1, 0.25, 0.5, 0.75$, using a multi-mass solver.

The Zolotarev approximation with 12 poles was used for the first 55
configurations. For the remaining 45 configurations, the Chebyshev
approximation (degree $100\sim 500$) was used, after Ritz projection of the
lowest 12 eigenvectors of $H^2$, as it was found to provide about 20\% better
performance. The convergence criterion used was $\vert D^\dag D \psi
-\chi\vert^2 <10^{-7}$.
     
The computations were performed with shared memory F90 code, optimized
and run on 16 and 32 processor IBM-p690 nodes at BU and NCSA.

\section{Meson Results}

\begin{figure}[t]
\vspace{-3mm}
\psfig{file=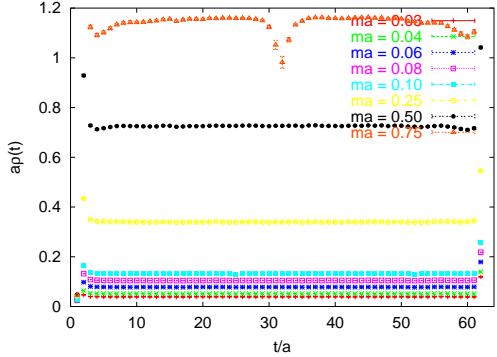,height=0.32\textheight,angle=-90}
\caption{AWI ratio as a function of time for all denerate quark mass
combinations.}
\label{fig:rho}
\end{figure}

\begin{figure}[t]
\vspace{-3mm}
\psfig{file=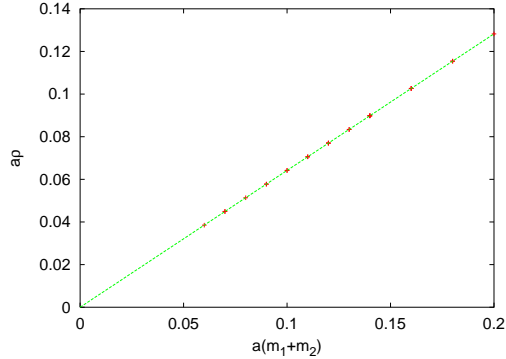,height=0.32\textheight,angle=-90}
\caption{Axial Ward identity, continued.}
\label{fig:awir}
\end{figure}

We begin with a verification of the axial Ward identity (AWI)
prediction for the ratio
\beq
\rho(t)=G_{\nabla_0 A_0 P}(t)/G_{P P}(t)
\eeq
as presented in Fig.~\ref{fig:rho}. The fit to 
$a\rho=A+a(m_{q1}+m_{q2})/Z_A+ C [a(m_{q1}+m_{q2})]^2$
gives $A = (3.09 \pm 9.57) \x 10^{-6}$,
$Z_A = 1.5570 \pm 0.0004$, and
$C = (-5.12 \pm 0.61) \x 10^{-3}$, as shown in Fig.~\ref{fig:awir}. 

\subsection{Meson Spectrum}

\begin{figure}[t]
\vspace{-3mm}
\psfig{file=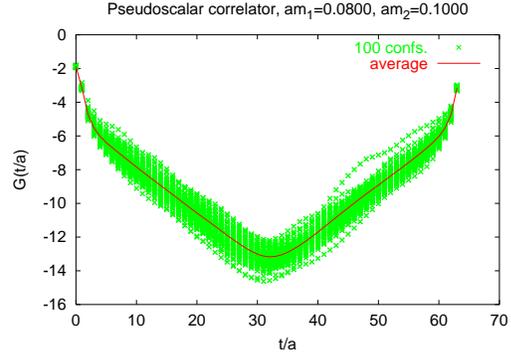,height=0.32\textheight,angle=-90}
\caption{Time dependence of a pseudoscalar correlator. Its value
  on each gauge configuration as well as its average value are shown.}
\label{fig:PPcorr}
\end{figure}

Pseudoscalar density 2-point functions were formed from quark
propagators and then fit to the usual cosh form to calculate the
ground state mass.
An example correlator is shown in Fig.~\ref{fig:PPcorr}, for input quark
masses $am_{q1}=0.08$ and $am_{q2}=0.10$.

\begin{figure}[t]
\vspace{-3mm}
\psfig{file=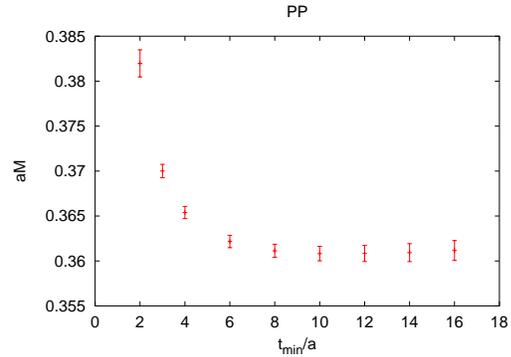,height=0.32\textheight,angle=-90}
\caption{Scanning the window for optimum $t_{min}$.}
\label{fig:tmin0810}
\end{figure}

\begin{figure}[t]
\vspace{-3mm}
\psfig{file=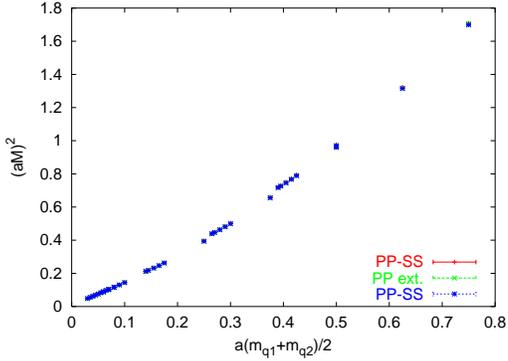,height=0.32\textheight,angle=-90}
\caption{Pseudoscalar spectrum.}
\label{fig:PPall}
\end{figure}

\begin{figure}[t]
\vspace{-3mm}
\psfig{file=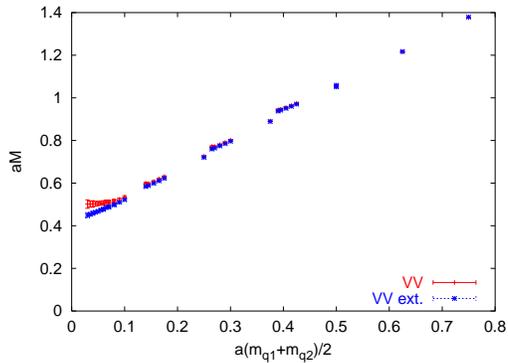,height=0.32\textheight,angle=-90}
\caption{Vector meson spectrum for point and extended sinks.}
\label{fig:VV}
\end{figure}

Effective mass plots are often used to determine the fitting window
$\{t_{min},t_{max}\}$ for the correlation function. In some cases
(e.g. for vector mesons), determining the best window can be
difficult. Therefore, we used a method that scanned the possible
windows, and then chose the smallest value of $t_{min}$ (consistent
with the errors) allowed before the clear effect of higher states
caused the mass prediction to rise. Use of a cosh fit allowed
$t_{max}$ to be fixed at 32.  An example of such a scan is shown in
Fig.~\ref{fig:tmin0810}.
The pseudoscalar spectrum for all possible input quark mass combinations is
shown in Fig.~\ref{fig:PPall}.
Using a larger lattice than \cite{Giusti:2001pk} allowed for
observation of vector meson states, as shown in Fig.~\ref{fig:VV}.

\subsection{Extended sink operators}

Since mesons are not point particles, an increased signal can be
obtained by using \i{extended} (or \i{smeared}) operators at the
source and sink~\cite{Hauswirth:2002eq}.
Matrix element calculations required point sources
and generating an additional set of propagators with extended
sources would have been computationally too costly.  Thus
we limited our analysis to extended sinks. 

\begin{figure}[t]
\vspace{-8mm}
\psfig{file=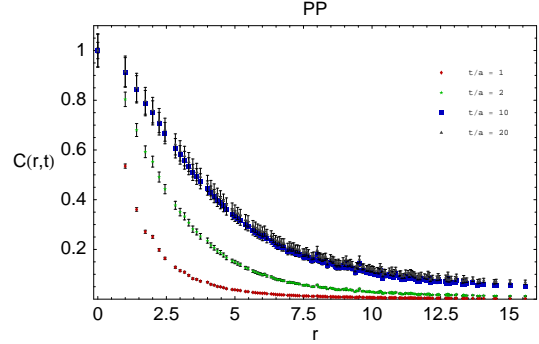,height=0.23\textheight}
\caption{Extended sink functions $C(r,t)$ for various $t/a$.}
\label{fig:PPcrt0303}
\end{figure}

The use of extended sinks allowed for a larger fitting window and
improved the prediction for mesonic observables. The extended sink
correlation function is
\beq
G_f(t)=
\langle \sum_{\bf x, y} \bar\psi(x) \Gamma \psi(y) 
f({\bf x-y}) \bar\psi(0) \Gamma \psi(0) \rangle
\eeq
where $x=({\bf x},t)$ and $y=({\bf y},t)$, and the gauge is fixed to
Landau gauge. 
With $f({\bf x-y})=\delta_r(|{\bf x -y}| - r)$,
we calculate the quantity $C(r,t)=G_{\d_r}(t)$ 
shown in Fig.~\ref{fig:PPcrt0303}. We then use the
asymptotic value $C_0(r) \defined C(r,\infty)$
to define an extended sink correlation function $G_{ext}(t) \defined
G_{C_0}(t)$. The functions $C(r,t)$ stabilizes at approximately
$t/a=10$, representing the meson final state ``wave function.''

\begin{figure}[t]
\vspace{-8mm}
\psfig{file=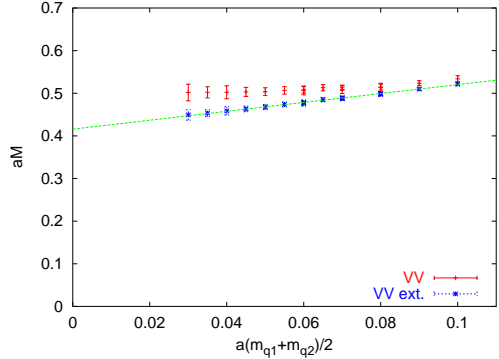,height=0.32\textheight,angle=-90}
\caption{Vector meson spectrum for $am_q\le 0.1$.}
\label{fig:VVlowmq}
\end{figure}

The use of extended sinks was most valuable in the calculation of the
vector meson spectrum, due to the increase in the size of the fitting
window. The point sink and extended sink vector meson spectra are
compared in Fig.~\ref{fig:VVlowmq} for low quark mass, $am_q\le 0.1$.
A fit of the quark-mass dependence of the vector meson mass obtained
from the extended sink data gives
\beq
aM=0.4159(11)+1.044(16)a(m_{q1}+m_{q2})/2\ .
\eeq

\subsection{Quenched chiral logs}

\begin{figure}[t]
\vspace{-3mm}
\psfig{file=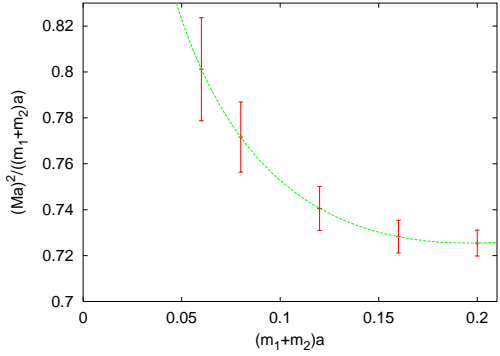,height=0.32\textheight,angle=-90}
\caption{Evidence for quenched chiral logs.}
\label{fig:msq1-PP-SS}
\end{figure}

The sharp rise at low $am_q$ of the ratio of the pseudoscalar mass
squared $(aM)^2$ to $(am_{q1}+am_{q2})$, which we observe in
Fig.~\ref{fig:msq1-PP-SS}, can be interpreted as an indication of
quenched chiral logs.  A value for the quenched chiral log parameter
$\d$ was obtained via a fit to the expression
$y=1+\d\,x$~\cite{Aoki:2002fd}, where
\beq
y=\fr{2m_1}{m_1+m_2}\fr{M_{12}^2}{M_{11}^2} \x
\fr{2m_2}{m_1+m_2}\fr{M_{12}^2}{M_{22}^2}
\eeq
and
\beq
x=2+\fr{m_1+m_2}{m_1-m_2}\,\ln\left(\fr{m_2}{m_1}\right)\ .
\eeq
The resulting value for $\d$ in the PP-SS channel is $\d=0.13 \pm 0.09$.

We fit the degenerate quark mass PP-SS channel results to the expression 
\beq
(Ma)^2=A(ma)^{1/(1+\d)}+B(ma)^2
\label{eq:MaSqQcl}
\eeq
where $M$ is the PP-SS mass and $m=m_{q1}+m_{q2}$. The fit is shown in
Fig.~\ref{fig:msq1-PP-SS}, and the results for the parameters are
\beq
A=0.441(1),\quad B=0.604(3),\quad \d=0.243(1),
\eeq
consistent with the value of $\d$ obtained above as well as
elsewhere~\cite{Aoki:2002fd,Wittig:2002ux,Gattringer:2003qx}.
Note that a fit to $(Ma)^2$ in Eq.~(\ref{eq:MaSqQcl}) containing an additional
constant term $C$ does not change the central values of $A$, $B$, or
$\d$ and produces a value for $C$ that is very small and consistent with zero.

\section{Baryon Results}

In calculating the baryon spectrum, we found that caution is needed
in determining an appropriate fitting window for the correlation
functions, due to fluctuations at large $t$.

\begin{figure}[t]
\vspace{-3mm}
\psfig{file=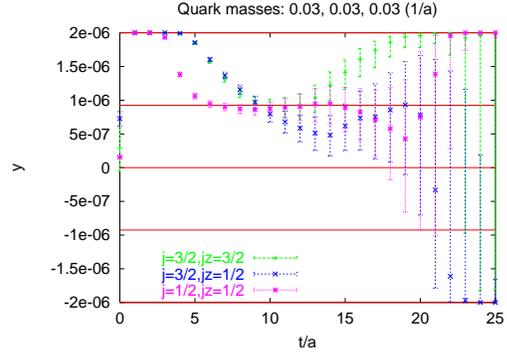,height=0.32\textheight,angle=-90}
\caption{Caution about fluctuations.}
\label{fig:m0303}
\end{figure}

Shown in Fig.~\ref{fig:m0303} are the average baryon propagators for
the octet and decuplet, after multiplication
by $\exp(0.65 t/a)$ and the mapping $ y \to 2 d \tanh [y/(2d)],
\;\;d=10^{-6}$. The mapping allows for a clear exposition of the fluctuations.

\begin{figure}[t]
\vspace{-3mm}
\psfig{file=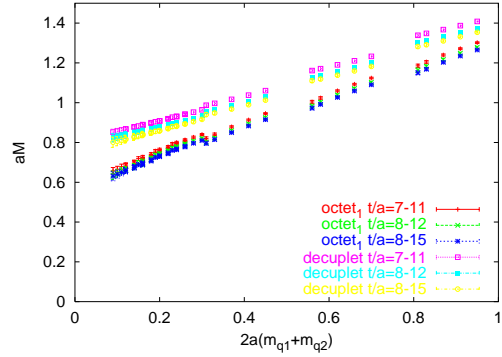,height=0.32\textheight,angle=-90}
\caption{Baryons: octet and decuplet spectra for various fitting windows.}
\label{fig:baryons1}
\end{figure}

Fig.~\ref{fig:baryons1} shows the octet and decuplet baryon masses vs.~total
quark mass for all quark mass combinations where two of the quarks have the 
same mass $m_{q1}$, while the third quark has mass $m_{q2}$, 
and for various correlator fitting windows. The sensitivity to the
fitting window is apparent. The ultimate fitting window chosen was $ 8
\le t/a \le 15$.

\begin{figure}[t]
\vspace{-3mm}
\psfig{file=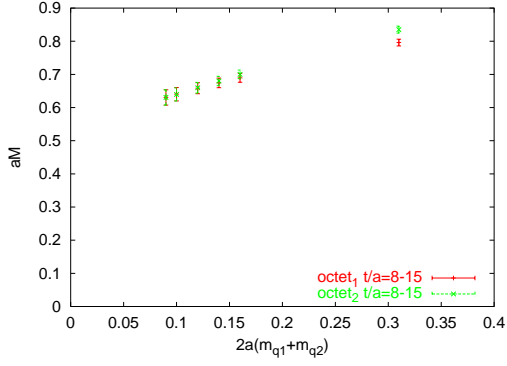,height=0.32\textheight,angle=-90}
\caption{Spectra for the two octet baryon states discussed in the text.}
\label{fig:o1o2}
\end{figure}

Fig.~\ref{fig:o1o2} shows the low quark mass spectrum, with fitting
window $8\le t/a \le 15$, for the two octet states
\beq
 8_1=(\vert \uparrow \downarrow \uparrow \;\rangle
-\vert \downarrow \uparrow  \uparrow \; \rangle)/\sqrt 2
\eeq
\beq
8_2=(\vert \uparrow \downarrow \uparrow \;\rangle
+\vert \downarrow \uparrow  \uparrow \; \rangle
-2 \vert \uparrow  \uparrow \downarrow \; \rangle)/\sqrt 6
\eeq
with $am_{q1} = 0.03$ and $0.03 \le am_{q2} \le 0.25$.
The corresponding correlation functions become identical for
$m_{q2}=m_{q1}$, but would describe $\Lambda$, $\Sigma$-like states,
respectively, for $m_{q1} \ne m_{q2}$. 

\begin{figure}[t]
\vspace{-3mm}
\psfig{file=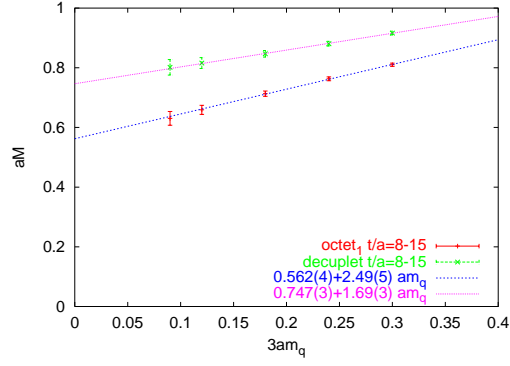,height=0.32\textheight,angle=-90}
\caption{Baryons for $am_{q_1}=am_{q_2}\le 0.1$.}
\label{fig:bdiag}
\end{figure}

Fig.~\ref{fig:bdiag} shows the degenerate quark mass octet and
decuplet spectrum for low quark mass ($am_q \le 0.1$) again with
fitting window $8\le t/a \le 15$. Extrapolating
to the chiral limit we get $aM_8=0.562(4)$, $aM_{10}=0.747(3)$, 
$aM_\rho=0.4159(11)$, and correspondingly
\beq
M_\rho/M_8=0.740(7)
\eeq
\beq
M_{10}/M_8=1.329(11)
\eeq
while experimentally $M_\rho/M_N=0.820$ and $M_\Delta/M_N=1.312$.

\section{Conclusion}
Our results show that use of the overlap formulation is practical,
albeit computationally costly, for the calculation of quark
propagators on large lattices, and validate the good chiral properties
of this discretization.  Work in progress includes the evaluation of
quark correlation functions in baryon final states and quenched
calculations on a $14^3 \x 48$ lattice at $\b=5.85$ to investigate
scaling.

\end{document}

%% file: macros.tex
\def\ben{\begin{equation}}
\def\een{\end{equation}}

\def\beq{\begin{equation}}
\def\eeq{\end{equation}}

\def\benn{\begin{equation*}}
\def\eenn{\end{equation*}}

\def\beqa{\begin{eqnarray}}
\def\eeqa{\end{eqnarray}}

\def\bea{\begin{eqnarray*}}
\def\eea{\end{eqnarray*}}

\def\bitem{\begin{itemize}}
\def\eitem{\end{itemize}}

\def\benum{\begin{enumerate}}
\def\eenum{\end{enumerate}}

\def\btabu{\begin{tabular}}
\def\etabu{\end{tabular}}

\def\btab{\begin{table}}
\def\etab{\end{table}}

\def\bfig{\begin{figure}}
\def\efig{\end{figure}}

\def\barr{\begin{array}}
\def\earr{\end{array}}

\def\bverb{\begin{verbatim}}
\def\everb{\end{verbatim}}

\def\defined{\equiv}

\def\fr#1#2{\,\frac{#1}{#2}\,}

\def\x{\times}

\def\approx{\simeq}

\def\i{\emph}

\def\b{\beta}
\def\d{\delta}

\newcommand{\rr}[1]{\let\temp=\\\raggedright#1\let\\=\temp}


%% file: spec_lat04.bbl
\begin{thebibliography}{99}

\bibitem{Berruto:2003rt}
F.~Berruto, N.~Garron, C.~Hoelbling, L.~Lellouch, C.~Rebbi and N.~Shoresh,
Nucl.\ Phys.\ Proc.\ Suppl.\  {\bf 129-130}, 471 (2004)
[arXiv:hep-lat/0310006].

\bibitem{lat04_wme}
F.~Berruto et~al., presentation by L.~Lellouch at this conference.

\bibitem{Giusti:2001pk}
L.~Giusti, C.~Hoelbling and C.~Rebbi,
Phys.\ Rev.\ D {\bf 64}, 114508 (2001)
[Erratum-ibid.\ D {\bf 65}, 079903 (2002)]
[arXiv:hep-lat/0108007].

\bibitem{Hauswirth:2002eq}
S.~Hauswirth,
arXiv:hep-lat/0204015.

\bibitem{Aoki:2002fd}
S.~Aoki {\it et al.}  [CP-PACS Collaboration],
Phys.\ Rev.\ D {\bf 67}, 034503 (2003)
[arXiv:hep-lat/0206009].

\bibitem{Wittig:2002ux}
H.~Wittig,
Nucl.\ Phys.\ Proc.\ Suppl.\  {\bf 119}, 59 (2003)
[arXiv:hep-lat/0210025].

\bibitem{Gattringer:2003qx}
C.~Gattringer {\it et al.}  [BGR Collaboration],
Nucl.\ Phys.\ B {\bf 677}, 3 (2004)
[arXiv:hep-lat/0307013].


\end{thebibliography}
